\documentclass[twocolumn,showpacs,preprintnumbers,amsmath,amssymb,prb]{revtex4-1}
\usepackage{graphicx}
\usepackage{dcolumn}
\usepackage[utf8]{inputenc}
\usepackage[ngerman, english]{babel}
\usepackage{hyperref}
\usepackage{bm}
\usepackage{xcolor}
\usepackage{mathtools}
\usepackage{tikz}
\newcommand{\avg}[1]{\left\langle{#1}\right\rangle}

\newcommand{\chat}[1]{\ensuremath{\xop{#1}}}
\newcommand{\tempop}[3][\textstyle]{\settowidth{\dimen1}{$#1\hat{#2}$}\makebox[\dimen1][l]{$#1\hat{#2\mspace{#3}}$}}
\newcommand{\xop}[1]{{\mathchoice{\tempop[\displaystyle]{#1}{3.5mu}}{\tempop{#1}{3.5mu}}{\tempop[\scriptstyle]{#1}{3.5mu}}{\tempop[\scriptscriptstyle]{#1}{3mu}}}}
\newcommand{\cbar}[1]{\ensuremath{\xbar{#1}}}
\newcommand{\cbarbar}[1]{\ensuremath{\xbarbar{#1}}}
\newcommand{\tempbarbar}[3][\textstyle]{\settowidth{\dimen1}{$#1\bar{\bar{#2}}$}\makebox[\dimen1][l]{$#1\bar{\bar{#2\mspace{#3}}}$}}
\newcommand{\tempbar}[3][\textstyle]{\settowidth{\dimen1}{$#1\bar{#2}$}\makebox[\dimen1][l]{$#1\bar{#2\mspace{#3}}$}}
\newcommand{\xbarbar}[1]{{\mathchoice{\tempbarbar[\displaystyle]{#1}{3.5mu}}{\tempbarbar{#1}{3.5mu}}{\tempbarbar[\scriptstyle]{#1}{3.5mu}}{\tempbarbar[\scriptscriptstyle]{#1}{3mu}}}} 
\newcommand{\xbar}[1]{{\mathchoice{\tempbar[\displaystyle]{#1}{3.5mu}}{\tempbar{#1}{3.5mu}}{\tempbar[\scriptstyle]{#1}{3.5mu}}{\tempbar[\scriptscriptstyle]{#1}{3mu}}}} 
\renewcommand{\d}{{\mathrm{d}}}

\newcommand{\lowerbossphantom}{\vphantom{\cbarbar{x}}}
\newcommand{\upperbossphantom}{\vphantom{\dagger}}
\newcommand{\aop}[2]{\ensuremath{\chat{c}_{#1,#2\lowerbossphantom}^{\upperbossphantom}}}
\newcommand{\cop}[2]{\ensuremath{\chat{c}_{#1,#2\lowerbossphantom}^{\dagger\upperbossphantom}}}

\newcommand{\kronecker}[2]{\delta^{\upperbossphantom}_{{#1},{#2}\lowerbossphantom}}
\newcommand{\sx}{s}

\renewcommand{\sp}{{\sx}^{\prime}}

\newcommand{\Sx}{\sigma}

\newcommand{\reffig}[1]{Fig.~\ref{#1}}

\begin{document}
\bibliographystyle{apsrev}

\title{Comment on ``On the unphysical solutions of the Kadanoff--Baym equations in linear response: Correlation-induced homogeneous density-distribution and attractors''}

\author{\begin{otherlanguage}{ngerman}N.~Schlünzen\end{otherlanguage}}
\author{J.-P.~Joost}
\author{M.~Bonitz}
\affiliation{\begin{otherlanguage}{ngerman}Institut für Theoretische Physik und Astrophysik, Christian-Albrechts-Universität zu Kiel, D-24098 Kiel, Germany\end{otherlanguage}}


\date{\today}
 
\begin{abstract}
In a recent Rapid Communication [A.~Stan, Phys. Rev. B \textbf{93}, 041103(R) (2016)], the reliability of the Keldysh--Kadanoff--Baym equations (KBE) using correlated selfenergy approximations applied to linear and nonlinear response has been questioned. In particular, the existence of a universal attractor has been predicted that would drive the dynamics of any correlated system towards an unphysical homogeneous density distribution regardless of the system type, the interaction and the many-body approximation. Moreover, it was conjectured that even the mean-field dynamics would be damped. Here, by performing accurate solutions of the KBE for situations studied in that paper, we prove these claims wrong, being caused by numerical inaccuracies.


\end{abstract}
\maketitle
The dynamics of correlated quantum many-body systems has been in the focus of experimental and theoretical studies over the recent two decades. Applications span (but are not limited to) nuclear physics, semiconductor optics and transport, dense plasmas and, more recently, strongly correlated materials and ultracold atoms \cite{pngf6}. A very popular tool to describe these systems theoretically has been the method of Nonequilibrium Green functions (NEGF) \cite{book_kadanoffbaym_qsm, keldysh} due to their internal consistency and conserving properties. For recent text book discussions see Refs.~\onlinecite{haug-jauho-book, bonitz-book, Robertbook, balzer-book}. Direct numerical solutions of their equations of motion---the Keldysh--Kadanoff--Baym equations (KBE)---have been performed for macroscopic, spatially homogeneous systems such as nuclear matter \cite{danielewicz}, dense plasmas and electron-hole plasmas (e.g. Refs.~\onlinecite{koehler96, kwong_pss98}), or the correlated electron gas \cite{kwong_prl00}. More recently, finite spatially inhomogeneous systems were treated, including atoms and small molecules \cite{dahlen_prl, balzer_pra10, balzer_pra10_2}, electrons in quantum dots \cite{balzer_prb09} or finite Hubbard clusters \cite{puigvonfriesen09, puigvonfriesen10, bonitz_cpp15}. For an overview see Ref.~\onlinecite{balzer-book}. 

Given the high success of numerical solutions of the KBE, which includes excellent agreement with time-resolved optical experiments in semiconductor optics, excitonic features and transport \cite{haug-jauho-book} and, recently, with experiments on the expansion dynamics of fermionic atoms \cite{schluenzen_prb16,schluenzen_cpp16}, it came as a surprise when unphysical behaviors were reported in applications to small systems. 
Von~Friesen, Verdozzi and Almbladh demonstrated~\cite{puigvonfriesen09,puigvonfriesen10} that, in small Hubbard clusters, cf. Eq.~(\ref{eq:hubbard}), subjected to a strong external potential, the nonlinear density evolution suffers from an unphysical damping, eventually leading to a steady state, in striking contrast to the exact solution. 
The authors explained this behavior by the highly nonlinear structure of the correlation selfenergies entering the KBE giving rise 
to an infinite sum of diagrams during a self-consistent solution of the KBE. Due to the partial summation schemes of the many-body approximations, the order-by-order balance of the exact solution can be violated which leads to an artificial energy reservoir that can cause damping. 
This explanation was supported by modified approximations where the degree of selfconsistency was reduced\cite{puigvonfriesen10}. Another confirmation and, at the same time, a more systematic approach to this problem is the application of the generalized Kadanoff--Baym ansatz (GKBA)\cite{lipavski} that practically eliminates the artificial damping\cite{hermanns_prb14}.\\

In view of the importance and popularity of the KBE, a detailed investigation of the issue of unphysical solutions and a clear mapping out of the range of validity of the KBE is, of course, of high interest. Such an analysis has been attempted by Stan~\cite{stan} who concludes that unphysical solutions are universal when solving the KBE with a correlation selfenergy, thereby ``[\ldots]drastically restricting the parameter space for which the method can give physically meaningful insights.''. It is the purpose of this Comment to analyze these far-reaching statements.

The author of Ref.~\onlinecite{stan} considers a one-band Hubbard model with the Hamiltonian\cite{error1}
\begin{align}
 H(t) &= - \sum_{\avg{\sx,\sp}}\sum_{\sigma=\uparrow,\downarrow}\cop{\sx}{\Sx}\aop{\sp}{\Sx} + U \sum_{\sx}\cop{\sx}{\uparrow}\aop{\sx}{\uparrow}\cop{\sx}{\downarrow}\aop{\sx}{\downarrow} \nonumber\\
 &\quad\, + \sum_{\sx}\sum_{\sigma=\uparrow,\downarrow} f_{\sx}(t) \cop{\sx}{\Sx}\aop{\sx}{\Sx} \,,
 \label{eq:hubbard}
\end{align}
with $\avg{\sx,\sp}$ being the summation over next neighbors and $U$ being the on-site Hubbard interaction. As a second example, he considers a Hubbard lattice  with Coulomb interaction. The analysis focuses on a simple system: two lattice sites occupied by two electrons (Hubbard dimer), except for one case where a four-site system is simulated. Furthermore, the interaction strength $U$ (in units of the hopping rate) is varied between $0$ and $5$ and the system is treated using weak coupling many-body approximations: the second Born selfenergy (2B, except for one case where also $GW$ results are shown). To study the electron dynamics following an external excitation, the author considers two variants of the time-dependent single-particle field $f_{\sx}(t)$: first, a step-like form, $f_{\sx}(t) = w_0 \kronecker{\sx}{1} \theta(t)$, and, second, an instantaneous excitation: $f_{\sx}(t)= k_0 \kronecker{\sx}{1} \delta (t)$, both acting only on site 1. 
Varying the field amplitude between $0.01$ and $5$ the linear and nonlinear response are investigated.
\\
Based on the simulation results for this limited set of systems and situations, the author draws the following conclusions that are termed ``universal'', i.e., are claimed to be valid {\it regardless of the system size, the interaction type, the interaction strength and the many-body approximation}:
\begin{enumerate}
 \item The density dynamics obtained from the KBE in the case of strong excitation is damped, in agreement with previous studies\cite{puigvonfriesen09, puigvonfriesen10}.
 \item For sufficiently long propagation time, a state with homogeneous density distribution (HDD) is reached, indicating the existence of an attractor. 
 \item In addition to previous observations, the unphysical damping occurs also for weak excitation (linear response regime).  
 \item For an uncorrelated system (Hartree or Hartree--Fock selfenergies), damping occurs as well, although no HDD is approached.
\end{enumerate}
We underline that item 1 is relevant only for small finite systems, i.e., the damping effect vanishes quickly with increasing system size. 
According to the author of Ref.~\onlinecite{stan}, the reason why the new points 2.--4. have been ``missed'' by previous studies is due to the insufficient propagation durations in the latter.
In the remainder of this Comment, we carefully test the above new claims for several relevant cases.

\begin{figure}
\includegraphics{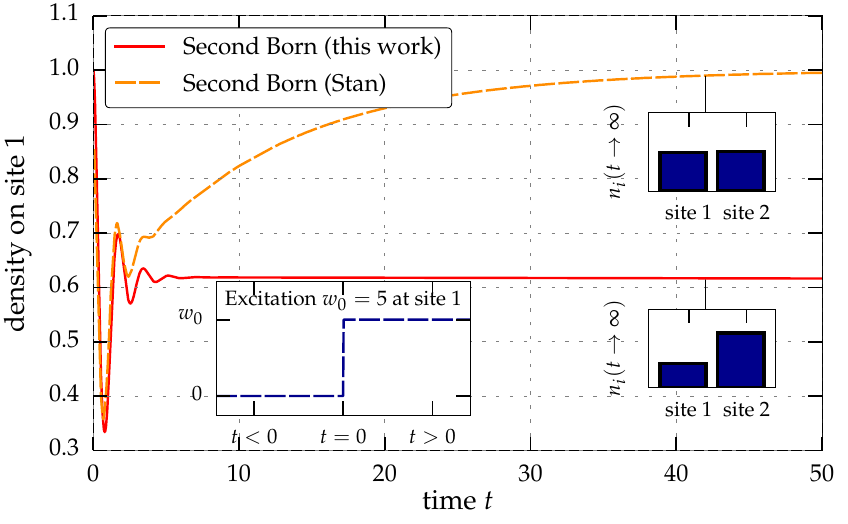}
 \caption{Density evolution on the first Hubbard site of the dimer with $U=4$ after the switch-on of a constant excitation with $w_0=5$ at site $1$, which is shown in the lower left inset. The insets on the right-hand-side show the asymptotic density distributions of Ref.~\onlinecite{stan} (top) and the present work (bottom). Solutions of the KBE in selfconsistent second Born approximation. The time step in our simulation is $\Delta t=10^{-3}$.}
 \label{fig:nonlinear}
\end{figure}

Let us start with item 2. and analyze the results presented in Fig.~1 of Ref.~\onlinecite{stan}. There the author studies the nonlinear response of a correlated dimer ($U=4$) to a strong step-like excitation ($w_0=5$). His result for the density on site 1 is reprinted in our Fig.~\ref{fig:nonlinear} by the dashed line \cite{units} indicating that the density approaches unity (the same value as on the other site, cf. upper inset), i.e., the dynamics approach a spatially homogeneous state (HDD). Now, compare this to our result\cite{method} shown by the full line. Both simulations are in agreement for short times, $t \lesssim 2$, after which we observe a qualitatively different behavior. Even though we also find the unphysical damping known from Refs.~\onlinecite{puigvonfriesen09, puigvonfriesen10}, the asymptotic value is very different from the one of Stan. Regardless of how far the simulations are continued, no HDD state emerges. We note that the time step in our simulations is $\Delta t = 10^{-3}$ whereas Stan reports the value~\cite{stan} $\Delta t \simeq 10^{-2}$. 
[a precise value for the time step is missing from his paper]. 
We underline that this is a typical case.
In converged simulations we never found a homogeneous density.

\begin{figure}
 \includegraphics{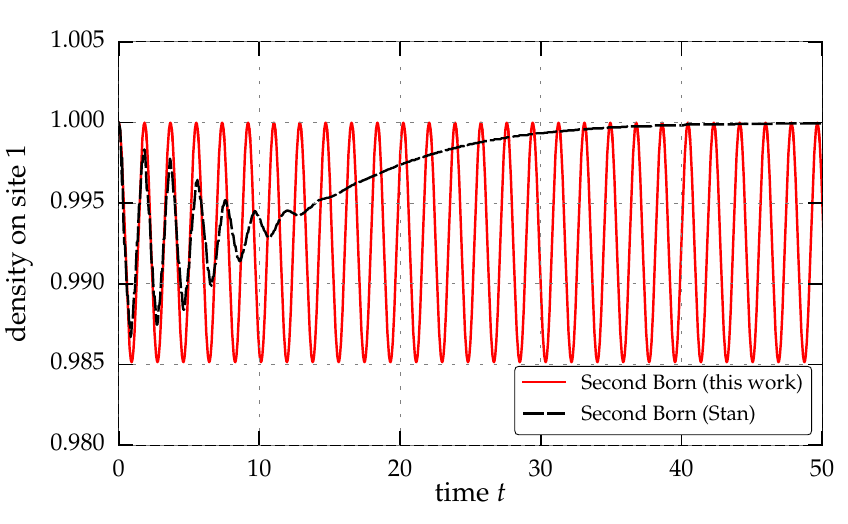}
 \caption{Density evolution at site 1 of a Hubbard dimer ($U=3$), following a very weak ($w_0=0.05$) step-like excitation at site 1. Black dashed line: result of Ref.~\onlinecite{stan}. Full red line: present result, using a time step of $\Delta t=10^{-3}$.}
 \label{fig:linear}
\end{figure}

Item 3. concerns the case of a very weak external excitation (linear response). Results for a two-site system were presented in Fig.~2 of Ref.~\onlinecite{stan}. Here, we concentrate on the example of a Hubbard system at $U=3$ excited by a weak external field (amplitude $w_0=0.05$) that is turned on at time $t=0$ at site 1. While the exact dynamics show undamped oscillations (cf. Fig.~2 (a) and (b) of Ref.~\onlinecite{stan}), Stan's second order Born result for the density at site 1 shows strong damping initially and, after $t\sim 10$, approaches the homogeneous density value $n=1$, cf. the black dashed curve in Fig.~\ref{fig:linear}. Our result is shown by the full red line and shows undamped oscillations as the exact solution.
We note that the amplitude and frequency of our result show small deviations from the exact data 
which is a consequence of the failure of the second Born approximation for $U$ exceeding unity \cite{hermanns_prb14,schluenzen_prb16}.

\begin{figure}
\includegraphics{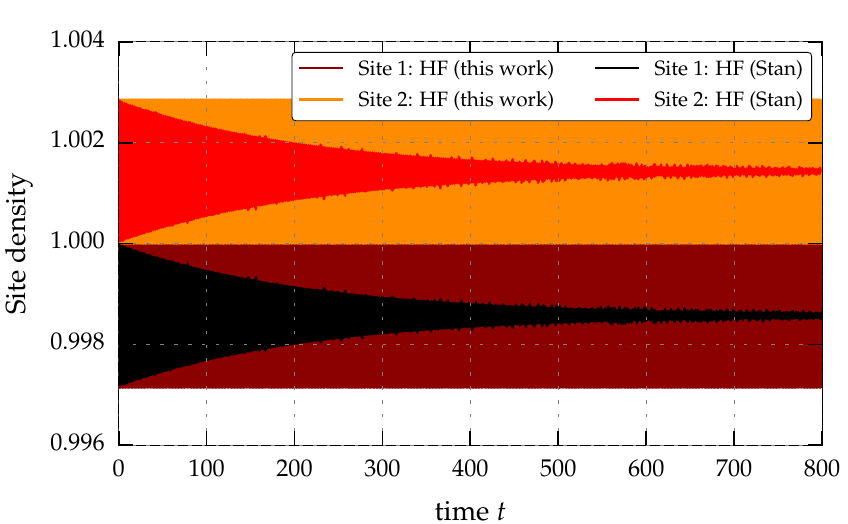}
 \caption{Mean field (Hartree) density evolution of a Hubbard dimer with $U=5$ following the switch-on of a constant excitation with $w_0=0.01$ on site $1$. The results of 
Ref.~\onlinecite{stan} are shown by the red and black lines and exhibit damping, whereas our results are undamped 
(orange and brown lines). Note the high-frequency oscillations of the density.}
 \label{fig:HF}
\end{figure}

Let us now turn to item 4. of the above list, which concerns the mean-field dynamics.
In Fig.~5 of Ref.~\onlinecite{stan}, a strongly interacting ($U=5$) dimer is considered 
in Hartree and Hartree--Fock (HF) approximations. The corresponding results of Stan for the densities on the two sites are reproduced in \reffig{fig:HF} (cf.  the red and black curves) and exhibit a damping towards constant (slightly different) values. This relaxation behavior is very surprising since mean-field dynamics are non-dissipative\cite{kbe-tdhf}.
We, therefore, repeated the Hartree simulations with our code for the same parameters. The results are plotted by the orange and brown curves and show no damping. 
We also note that in our simulations the density exhibits high-frequency oscillations. These oscillations are also present in the data of Stan but their frequency is substantially lower than ours. 
%

%
\begin{figure}
 \includegraphics{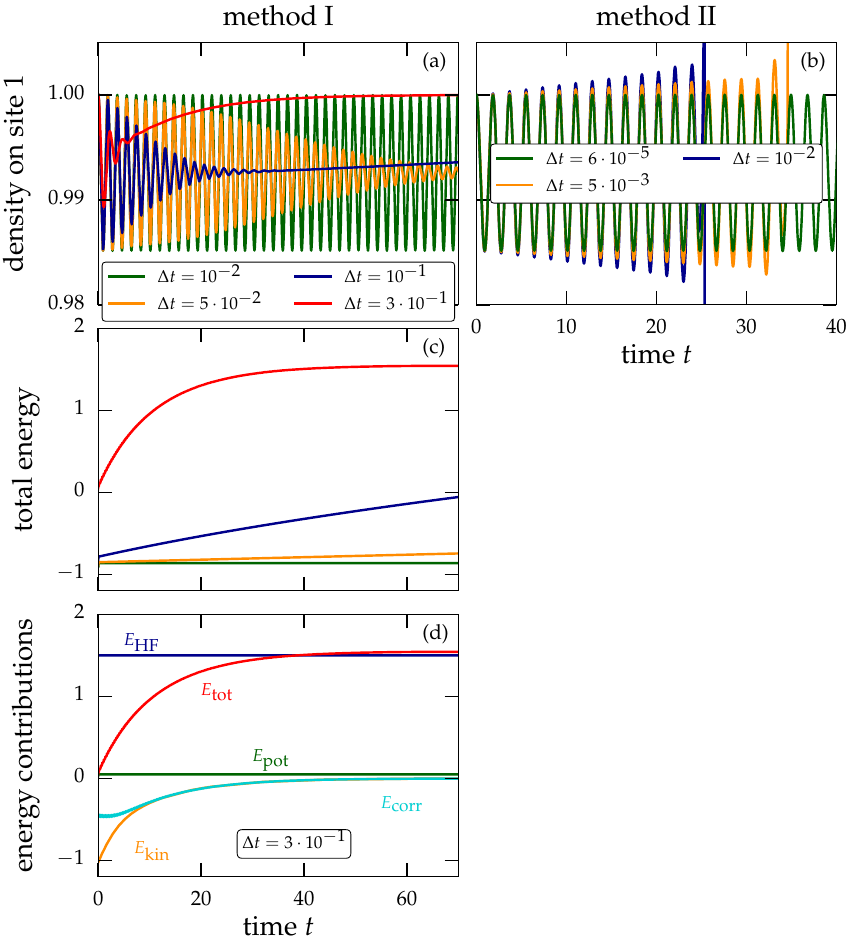}
 \caption{Demonstration of the convergence behavior for a Hubbard dimer ($U=3$), following a very weak ($w_0=0.05$) step-like excitation at site 1. (a)/(b): density evolution at site 1 for different time steps $\Delta t$. Simulations in Fig.~(a) (Fig.~(b)) are done with method I (method II). (c): quality of energy conservation corresponding to the results in (a). (d): different energy contributions for $\Delta t = 3\cdot 10^{-1}$ in (a) and (c). We note that in (b) the time steps only refer to the integration (B). The collision integral (A) is solved with $\Delta t= 10^{-2}$.}
 \label{fig:convergence}
\end{figure}

Summarizing our numerical simulations (cf. Figs.~\ref{fig:nonlinear}--\ref{fig:HF}) we found that the statements 2.--4. of the above list cannot be reproduced within converged calculations. By ``converged'' we denote simulations the result of which does not change anymore upon further reduction of the time step in the discetization of the KBE. 
To understand possible sources of damping in the linear response regime and the emergence of an HDD state we now analyze the convergence behavior in detail. 
The numerical solution of the KBE basically invokes two time integration procedures\cite{stan_2009,schluenzen_cpp16}:
\begin{itemize}
\item [(A)] the evaluation of the collision integral (cf. integral expression in Eq.~(1) of Ref.~\onlinecite{stan}) and
\item [(B)] the time propagation of the entire (integro-)dif\-ferential equations (time-stepping).
\end{itemize}
Obviously, for any discretization procedure, the exact integro-differential equation will be recovered when the time step $\Delta t$ vanishes. For practical simulations, however, a finite value $\Delta t$ has to be used, so the question arises, which values are acceptable. For converged solutions, all values $\Delta t$ less or equal some threshold $\Delta t_c$ are expected to yield the same result, at least for a given propagation duration $T$ ($\Delta t_c$ may depend on $T$). A key question is how to determine the threshold $\Delta t_c$. Since the answer to these questions strongly depends on the specific scheme used to perform the integrations (A) and (B), we consider two typical cases:
\begin{itemize}
\item [(I)] The collision integral (A) is evaluated in the lowest possible order using the trapezoidal rule whereas the integration (B) is performed by a fourth order Runge--Kutta method.
\item [(II)] The integral evaluation (A) is performed using a higher order scheme (see Ref.~\onlinecite{schluenzen_cpp16} for details), and the integration (B) is done with an explicit Euler method which is known to be less accurate than Runge-Kutta.
\end{itemize}
In both cases convergence can be achieved, however, the threshold values $\Delta t_c$ maybe different.

In the following, we analyze these issues for the setup presented in Fig.~\ref{fig:linear} [i.e. a dimer ($U=3$) with a weak step-like excitation ($w_0=0.05$) at site 1], but the results are representative for all examples considered in this paper. 
Figure~\ref{fig:convergence} (a) shows the density evolution using method I and different time steps $\Delta t$ ranging from $\Delta t = 0.3$ to $\Delta t =0.01$. In (b) the convergence behavior for the density is shown for method II for time steps in the range $\Delta t = 6 \cdot 10^{-5} \dots 0.01$. In both cases convergence is observed: undamped density oscillations that are in exact agreement with each other (see also our result in Fig.~\ref{fig:linear}) and are depicted by the green curve.
Since the two implementations are independent of each other, this provides a strong test of the numerics. At the same time, both methods have a very different numerical efficiency that is reflected by the threshold time steps: in case of method I, $\Delta t^I_c \approx 0.01$, whereas for method II, $\Delta t^{II}_c \approx 6 \cdot 10^{-5}$. 

Let us now analyze the behavior of the simulations when the time step exceeds $\Delta t_c$. The figure clearly demonstrates that  then the dynamics strongly deviate from the converged behavior where the type of density response and of deviation from the converged result is very different for methods I and II.
In case II [Fig.~\ref{fig:convergence} (b)] not-converged simulations lead to an increase of the oscillation amplitude in time and, eventually, the simulations become unstable. Increasing the time step leads to an earlier onset of the instability and a more rapid density increase. In case of method I [Fig.~\ref{fig:convergence} (a)], we observe the opposite behavior, for $\Delta t > \Delta t^I_c$: the density rapidly decays (cf. the yellow and green curves), a trend that sets in earlier when $\Delta t$ increases. If $\Delta t$ is increased to $0.1$ or beyond, however, the behavior changes: after a short decay interval the density increases again and approaches a constant value $n_1=n_2=1$, i.e. we exactly recover the
 trends  reported by Stan in Ref.~\onlinecite{stan} and that he termed ``emergence of the HDD'' or of a ``universal attractor''.
 From the above observations, we conclude that, indeed, a HDD can be found, however, only if the time step significantly exceeds the critical time step and only for certain discretization schemes. 
 Therefore, this observation is clearly a consequence of non-converged simulation and is not an inherent property of the KBE.

One may now ask how such erroneous simulations can be  avoided. The final test is always a verification of convergence, i.e. a repetition of the simulations with systematic reduction of  the time step $\Delta t$. In case of the KBE, fortunately, this procedure may be simplified essentially by monitoring the conservation laws of density and total energy. While the former is usually well maintained, the latter is quickly violated if the time step is chosen too large. We, therefore, present in 
Fig.~\ref{fig:convergence} (c) the time dependence of total energy for method I, for different time steps  [the behavior is similar for method II]. While for $\Delta t \le \Delta t_c$ energy is perfectly conserved (green curve), for larger time steps this conservation is violated, and the deviations increase with $\Delta t$. Comparison with figure (a) clearly shows that 
 an occurrence of damping goes together with a crucial violation of energy conservation\cite{conservation}. We also observe that the emergence of the HDD is connected to a convergence of the total energy to an unphysical value (cf. red curve). This can be understood from the fact that the trapezoidal rule systematically underestimates the result of the integration of oscillating functions, such as the integrand of the collision integral (see Appendix~\ref{app} for details). Together with the selfconsistent structure of the KBE, this results in an ongoing damping, up to the point when the collision integral completely vanishes. This is explored in more detail in Fig.~\ref{fig:convergence} (d) where the different contributions to the energy are shown for the time step $\Delta t = 3 \cdot 10^{-1}$. The potential and the HF energy are stable since they only depend on the density which is conserved due to the accurate solution of the differential equation (B). However, the kinetic and correlation energy, which are connected to the collision integral, tend to zero, leaving the system in a completely uncorrelated stationary state that has nothing to do with the Hamiltonian. Thus, for practical purposes, monitoring total energy conservation is a strong quality test giving a necessary (though not sufficient) criterion of convergence.

Another useful test of the accuracy of the simulations is the verification of time reversal symmetry---a known property of the KBE. This can be done in two ways. First, if after a propagation duration $t_1$, the times are inverted, $t \to -t$, a numerically correct scheme will return to the initial state after a time $2 t_1$. This behavior was verified by Stan in the supplementary material to Ref.~\onlinecite{stan}, but this only proves that the time step for integrating the differential equation (B) is sufficiently small, but it is independent of the accuracy of evaluation of the collision integral (A), as we show in Appendix~\ref{app_2}. 
Therefore, a more sensitive approach to time reversal is to change, at time $t_1$, instead, the sign of the Hamiltonian, $H(t) \to -H(-t)$ and of all its contributions. Any converged solution will return to the initial state at $t=2t_1$. In contrast, in case of a non-converged evaluation of the collision integral (A), time reversal symmetry is violated (there is a loss of information). 
This is demonstrated in App.~\ref{app_2} where we also show that the damped dynamics in the case of strong excitation of a small system (a known property of the KBE, cf. Refs.~\onlinecite{puigvonfriesen09,puigvonfriesen10} are completely time reversible, if the simulation is converged.
%

Let us summarize our results. We have repeated a representative part of the simulations of Ref. \onlinecite{stan} and presented the results in Figs.~\ref{fig:nonlinear}--\ref{fig:HF}. Our results are in disagreement with Ref.~\onlinecite{stan} on all 
the above points, 2.--4. In particular, we do not observe a HDD state in any of our simulations.
Our results have been obtained by two independent methods (method I and II) and have also been confirmed by another program~\cite{uimonen_pc}. 
In the second part of the paper we have analyzed 
 possible reasons of the disagreement with Ref. \onlinecite{stan}. A detailed analysis of the  convergence behavior of numerical solutions of the KBE has been summarized in Figs.~\ref{fig:convergence}, \ref{fig:colint} and \ref{fig:timereversal}. We presented numerical evidence that our results are converged. In contrast, the author of Ref. \onlinecite{stan} did not present such evidence. The data for the density conservation and time reversal in that paper are not conclusive and the crucial checks of total energy conservation and convergence with respect to the time step are missing.
Finally, by analyzing various numerical schemes and their convergence properties we were, indeed, able to recover the emergence of a HDD state of Ref. \onlinecite{stan}, however, only if we use method I together with a substantially too large time step. Thus, the predictions of Stan are wrong, being a numerical artifact [most likely arising from an inaccurate time integration of the collision integral, cf. Fig.~\ref{fig:convergence}]. The  impressive properties of the Keldysh--Kadanoff--Baym equations   remain fully intact.

We thank A.-M.~Uimonen for independent numerical confirmation of our results.
We acknowledge stimulating discussions with S.~Hermanns, G.~Stefanucci, R.~van~Leeuwen, and C.~Verdozzi and financial support by the Deutsche Forschungsgemeinschaft via grant BO 1366/9.

\appendix

\section{\label{app}Details on the numerical error of the trapezoidal rule}
\begin{figure}
\includegraphics{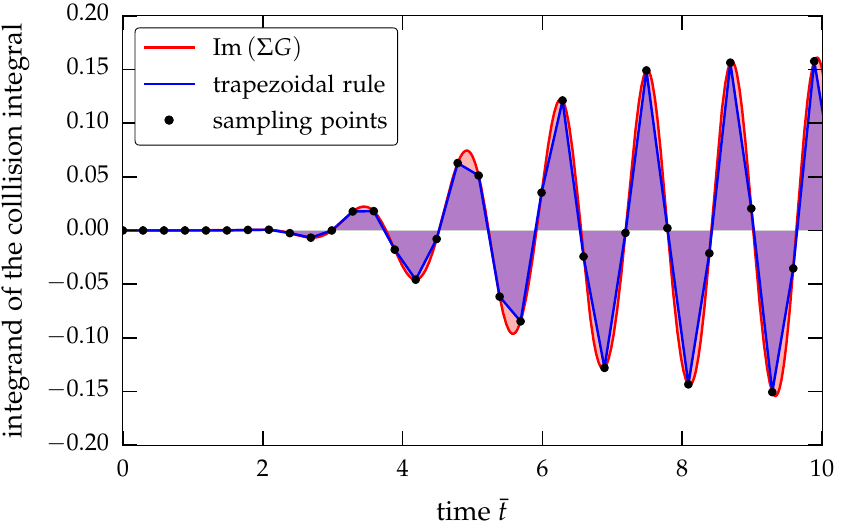}
 \caption{Illustration of the trapezoidal rule for a typical calculation of the collision integral. The red line shows a realistic example of the integrand in Eq.~(1) of Ref.~\onlinecite{stan} during a converged simulation. The blue curve corresponds to the respective approximation of the trapezoidal rule for a large time step $\Delta t = 0.3$.}
 \label{fig:colint}
\end{figure}

To understand the fact that numerical integration applying the trapezoidal rule can lead to an artificial damping in the solution of the KBE, it is instructive to look at the shape of the collision integral and its integrand, respectively. Fig.~\ref{fig:colint} shows a typical $\xbar t$-dependence of $\mathrm{Im}\left[ \Sigma \left(t,\xbar t \right) G \left( \xbar t, t' \right)\right]$ (red). As one can see the integrand oscillates around zero alternating between concave and convex pieces, depending on the sign. The blue line shows how the integrand is approximated with the trapezoidal rule integration. It is apparent that the absolute value of the integrand is systematically underestimated for every $\xbar t$. During the evaluation of the integral, after the cancellation of the areas with opposite sign, this leads to an underestimation of the collision integral. Due to the selfconsistent structure of the KBE this systematic numerical error results in a progressive damping during the time evolution eventually leading to vanishing  kinetic and correlation energy, cf. Fig~\ref{fig:convergence} (d). 

The damping property for the integration of oscillating functions can also be understood from a mathematical point of view. The error of the extended trapezoidal rule, $E(I)$, for the integral $\int_a^b f(x) \d x$ is given by\cite{atkinson}
\begin{eqnarray}
 E(I) = -\frac{h^2}{12} \left[f'(b) - f'(a) \right] + \mathcal{O}\left( h^3 \right)\, ,
\end{eqnarray}
where $h$ is the integration step. The behavior of oscillating integrands can be easily demonstrated for the example of a cosine function. For $I(x) = \int_0^x \cos\left(\cbar x\right) \d \xbar x$ it immediately follows that
\begin{eqnarray}
 I_\mathrm{trapez}(x) = \sin\left( x \right) \left[1-\frac{h^2}{12} \right] + \mathcal{O}\left(h^3 \right),
\end{eqnarray}
where the reduction of the amplitude is evident. 
We note that this systematic underestimation of the oscillations is inherent only for the trapezoidal rule. Higher order interpolation polynomials do not have this clear trend and, therefore, never result in an ``amplitude death''.

\section{\label{app_2} Time reversibility}
Beside the conservation of the particle number and the total energy, a very important accuracy test for the propagation of the KBE is provided by the time reversal symmetry. As mentioned in the main text, time reversal can be realized either by changing the direction of time, or by changing the sign of the Hamiltonian at some time $t_1$. 

\begin{figure}
\includegraphics{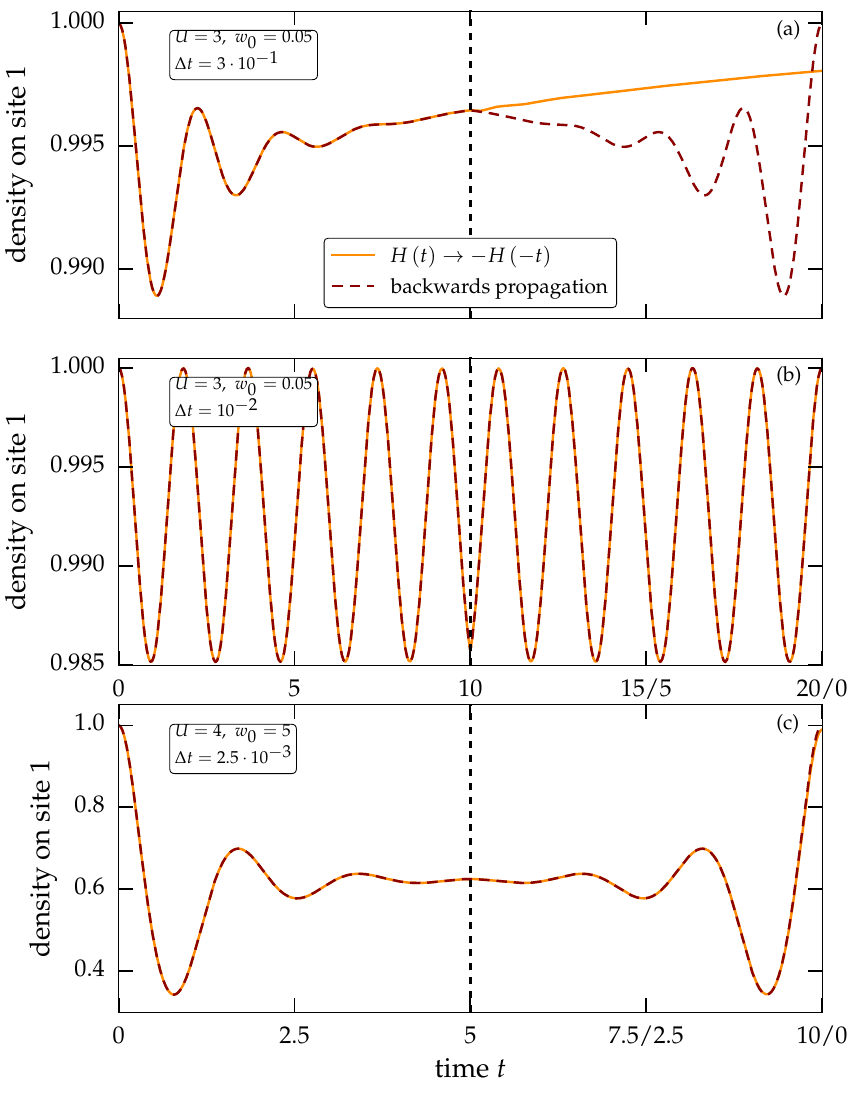}
 \caption{Time reversal properties of the density on the first site for a Hubbard dimer with step-like excitation. Solid yellow (dashed brown) lines correspond to simulations where $\hat H(t) \to -\hat H(-t)$ ($t \to -t$) is being applied. All calculations are performed via method I. Parts (a) and (b) show the density for linear response ($w_0 = 0.05$) and $U=3$. The time step in (a) is $\Delta t = 3 \cdot 10^{-1}$ (not converged), while in (b) it is $\Delta t = 10^{-2}$ (converged). Part (c) shows the time reversal behavior for a strong excitation ($w_0 = 5$) with $U=4$ and a time step of $\Delta t = 2.5 \cdot 10^{-3}$ (converged).}
 \label{fig:timereversal}
\end{figure}
In Fig.~\ref{fig:timereversal} (a) and (b) time reversibility tests are performed for linear response,  cf. Fig.~\ref{fig:linear} and Fig.~\ref{fig:convergence} (a).
In Fig.~\ref{fig:timereversal} (a) method I is used, with a time step of $\Delta t = 3\cdot 10^{-1}$ which was shown to result in a non-converged density evolution associated with damping and emergence of the HDD. While in the case of the backwards propagation ($t\to -t$, dashed brown curve), time reversal symmetry holds due to the accurate treatment of the time-stepping (B), this symmetry is completely broken if one applies the sign change of the Hamiltonian (solid yellow curve). This is a clear indication of a too large time step in the integral (A). In Fig.~\ref{fig:timereversal} (b) the behavior is shown for a converged calculation with $\Delta t = 10^{-2}$, resulting in an undamped density evolution. As expected, the results for both ways of performing the time reversal coincide and the system properly returns to the initial state.  

Finally, in Fig.~\ref{fig:timereversal} (c) we analyze the case of a strong excitation (${w_0=5}$), where unphysical damping of the density occurs in a converged solution (cf. Fig.~\ref{fig:nonlinear}). As one can see, even though the oscillation amplitude is drastically reduced, the propagation is entirely time reversal symmetric, even if the sign of the Hamiltonian is changed. Compared to Fig.~\ref{fig:timereversal} (a), this again confirms the substantial difference between the artificial damping for strongly excited systems (which is inherent to the KBE) and the damping caused by numerical inaccuracies.


\begin{thebibliography}{26}

\bibitem{pngf6} for a recent overview, see {\em Progress in Nonequilibrium Green's Functions VI}, J. Phys. Conf. Ser. vol. {\bf 696} (2016).

\bibitem{book_kadanoffbaym_qsm}{L.P.~Kadanoff and G.~Baym, \textit{Quantum Statistical Mechanics} (Benjamin, New York, 1962).}

\bibitem{keldysh} L.V.~Keldysh, ZhETF \b{47}, 1515 (1964).

\bibitem{haug-jauho-book} H. Haug, and A.-P. Jauho, {\em Quantum Kinetics in Transport and Optics of Semiconductors}, (Springer 2008).

\bibitem{bonitz-book} M. Bonitz, {\em Quantum Kinetic Theory} (Teubner 1998, 2nd ed. Springer 2016).

\bibitem{Robertbook} G.~Stefanucci, and R.~van Leeuwen, {\em Nonequilibrium many-body theory of quantum systems} (Cambridge University Press 2013).

\bibitem{balzer-book} K. Balzer, and M. Bonitz, ``Nonequilibrium Green's Functions Approach to Inhomogeneous Systems'',
Lect. Notes Phys. vol. {\bf 867} (2013).

\bibitem{danielewicz} P. Danielewicz, Ann. Phys. (N.Y.) {\bf 152}, 305 (1984).

\bibitem{koehler96} M. Bonitz, D. Kremp, D.C. Scott, R. Binder, W. D. Kraeft, and H. S. K\"ohler,
J. Phys. Condens. Matter {\bf 8}, 6057  (1996).


\bibitem{kwong_pss98} N.H. Kwong, M. Bonitz, R. Binder and S. K\"ohler,
phys. stat. sol. (b) {\bf 206}, 197 (1998). 

\bibitem{kwong_prl00}  N.H. Kwong, and M. Bonitz,
Phys. Rev. Lett. {\bf 84}, 1768 (2000). 

\bibitem{dahlen_prl}  N.E. Dahlen, and R. van Leeuwen, Phys. Rev. Lett. {\bf 98}, 153004 (2007).

\bibitem{balzer_pra10}  K. Balzer, S. Bauch, and M. Bonitz,
Phys. Rev. A {\bf 81}, 022510 (2010).

\bibitem{balzer_pra10_2}  K. Balzer, S. Bauch, and M. Bonitz,
Phys. Rev. A {\bf 82}, 033427 (2010).

\bibitem{balzer_prb09} K.~Balzer, M.~Bonitz, R.~van~Leeuwen, A.~Stan and N.~E.~Dahlen,
Phys. Rev. B {\bf 79}, 245306 (2009).



\bibitem{puigvonfriesen09}{M.~Puig von Friesen, C.~Verdozzi and C.-O.~Almbladh,
Phys.~Rev.~Lett.~\textbf{103}, 176404 (2009).}

\bibitem{puigvonfriesen10}{M.~Puig von Friesen, C.~Verdozzi and C.-O.~Almbladh,
Phys.~Rev.~B~\textbf{82}, 155108 (2010).}

\bibitem{bonitz_cpp15} M.~Bonitz, S. Hermanns, and N. Schl\"unzen,
Contrib. Plasma Phys. {\bf 55}, 152 (2015). 

\bibitem{schluenzen_prb16} N.~Schlünzen, S. Hermanns, M. Bonitz, and C. Verdozzi, Phys. Rev. {\bf 93}, 035107 (2016).

\bibitem{schluenzen_cpp16} N.~Schlünzen, and M. Bonitz, Contrib. Plasma Phys. {\bf 56}, 5 (2016).

\bibitem{lipavski} P.~Lipavsk{\'y}, V.~\ifmmode \check{S}\else \v{S}\fi{}pi\ifmmode \check{c}\else \v{c}\fi{}ka and B.~Velick{\'y}, 
Phys.~Rev.~B~\textbf{34}, 6933 (1986).

\bibitem{hermanns_prb14} S. Hermanns, N.~Schl\"unzen, and M. Bonitz, Phys. Rev. B {\bf 90}, 125111 (2014).



\bibitem{stan}{A.~Stan, Phys. Rev. B \textbf{93}, 041103(R) (2016).}

\bibitem{units} We use the common units for the density--the number of particles. This means, in all figures, our densities are larger by a factor of two, compared to Ref.~\onlinecite{stan}.

\bibitem{method} If not stated otherwise, our simulations are obtained using a fourth order Runge--Kutta scheme for the time-stepping (B) and a higher order integral evaluation (see Ref~\onlinecite{schluenzen_cpp16} for details) for the collision integral (A).

\bibitem{error1} The Hamiltonian in Ref.~\onlinecite{stan}, Eq.~(2), is wrong, missing the interaction matrix element.

\bibitem{kbe-tdhf} We underline that this has nothing to do with the Kadanoff--Baym equations because in the absence of a correlation selfenergy the KBE reduce to standard time-dependent Hartree--Fock.

\bibitem{stan_2009}{A.~Stan, N.~E.~Dahlen, and R.~van~Leeuwen, J.~Chem.~Phys {\bf 130}, 224101 (2009)}.

\bibitem{conservation} We note that the particle number is conserved within machine precision due to the accurate treatment of the time-stepping (B).

\bibitem{uimonen_pc} A.-M. Uimonen, private communication (2016).

\bibitem{atkinson}{K.~E.~Atkinson, \textit{An introduction to numerical analysis. Second edition} (Wiley, 1989).}



\end{thebibliography}
\end{document}